\documentclass{iau-JDSS}
\usepackage{graphicx}

\title[Complex asteroseismology of the B-type pulsators] 
{Complex asteroseismology of the B-type main sequence pulsators}

\author[J. Daszy\'nska-Daszkiewicz, P. Walczak]   
{Jadwiga Daszy\'nska-Daszkiewicz, Przemys{\l}aw Walczak}

\affiliation{Instytut Astronomiczny, Uniwersytet Wroc{\l}awski, ul. Kopernika 11, 51-622 Wroc{\l}aw, Poland
\break emails: daszynska@astro.uni.wroc.pl, walczak@astro.uni.wroc.pl}

\pubyear{2009}
\volume{Volume 15}  
\pagerange{20--21}
\date{?? and in revised form ??}
\jname{Highlights of Astronomy, Volume 15}
\editors{Ian F Corbett, ed.}
\begin{document}

\maketitle

\begin{abstract}
We present examples of an extended asteroseismic modelling in which we aim at fitting
not only pulsational frequencies but also certain complex parameter related to each
frequency. This kind of studies, called \textbf{complex asteroseismology},
has been successfully applied to a few main sequence B-type pulsators and
provided, in particular, plausible constraints on \textbf{stellar opacities}.
Here, we briefly describe our results for three early B-type stars.
\end{abstract}


\section{Introduction}
It is now well known that pulsational frequencies are a primary seismic probe
of stellar structure and evolution. A power of these seismic tools has been
firstly explored for the Sun and then for various types of pulsating variables
at different evolutionary stages.
Important results have been obtained also for the main sequence pulsators
of the B spectral type. Knowledge of the internal structure and evolution
of these massive stars is of great importance for astrophysics because
they form the CNO elements and are progenitors of Type II Supernovae.

Each pulsational frequency, $\nu$, is associated with the complex, nonadiabatic parameter $f$,
which describes the bolometric flux perturbation normalized
to the radial surface displacement. The $f$ parameter is embedded in the expression
for a complex amplitude of the light variations. Thus, having multicolour photometric
data, one can try to extract empirical values of $f$ together with the mode degree, $\ell$,
and compare them with theoretical counterparts.
A value of $f$ is determined in subphotospheric layers which have rather weak
contribution to the frequency value. Therefore, $\nu$ and $f$ constitute
the two asteroseismic tools complementary to each other and combining them
in a seismic survey yields a new kind of information.
In the case of the B-type pulsators, complex asteroseismology provides
a critical and unique test for stellar opacities and the atomic physics.

In this short report, we summarise our results for three early B-type pulsators:
$\theta$ Ophiuchi, $\nu$ Eridani and $\gamma$ Pegasi.

\section{The $\beta$ Cep star $\theta$ Ophiuchi}
$\theta$ Oph is the B2IV type star in which 7 pulsational frequencies were
detected from photometry in a range from 7 to 8 c/d. Three of them appeared
also in spectroscopy.
Using both the OP and OPAL opacity data, we found a family of seismic models with
different parameters ($M$, $T_{\rm eff}$, $\alpha_{ov}$, $Z$),
which reproduced two centroid frequencies: $\nu_3$ ($\ell=0,~p_1$) and $\nu_6$ ($\ell=1,~p_1$).
In general, seismic models with lower metallicity, $Z$, demanded a higher core overshooting, $\alpha_{ov}$.
Then, we went a step further by a requirement of fitting simultaneously also the $f$
parameter corresponding to the radial mode.
A comparison of empirical and theoretical values of $f$ pointed
substantially to a preference for the OPAL tables.
For more details see Daszy\'nska-Daszkiewicz \& Walczak (2009a).

\section{The hybrid  $\beta$ Cep/SPB pulsators: $\nu$ Eridani and $\gamma$ Pegasi}
$\nu$ Eri and $\gamma$ Peg are the most multimodal pulsating stars of early B spectral types.
Asteroseismic studies of these variables have been intensified
after the recent photometric and spectroscopic multisite campaigns which led
to a detection of next frequencies typical for the $\beta$ Cep type
as well as entirely new peaks in the SPB frequency domain.

The frequency analysis of the $\nu$ Eri data revealed 14 peaks: 12 of the $\beta$ Cep type
and 2 low frequency modes typical for the SPB pulsations.
In our seismic analysis, we first looked for stellar models with different ($M$, $T_{\rm eff}$, $\alpha_{ov}$, $Z$),
which reproduced three centroid frequencies:
$\nu_1$ ($\ell=0,~p_1$), $\nu_4$ ($\ell=0,~g_1$) and $\nu_6$ ($\ell=1,~p_1$).
Available data allowed us to determine the $f$ parameter for eight high frequency modes
and one SPB mode.
Then, we could compare empirical and theoretical values of $f$ in a wide range
of pulsational frequencies, i.e., from 0.6 to 8 c/d. The obtained consistency 
is very encouraging and brings a great seismic potential. The value of $f$ 
depends not only on mode frequency, but also on the shape of eigenfunctions.
In the low frequency region, $f$ strongly depends on the mode degree, $\ell$,
whereas for the high frequency modes it is independent of $\ell$.

A closer comparison of empirical and theoretical values of $f$ for the
radial mode indicated again a preference for the OPAL opacities.
Moreover, the OPAL seismic models had larger effective temperatures and masses
which fit better observational values of $T_{\rm eff}$ and luminosity.
More details can be found in Daszy\'nska-Daszkiewicz \& Walczak (2009b).

For more than 50 years, $\gamma$ Peg was considered as a monoperiodic star.
The analysis of the recent MOST and ground based photometric and spectroscopic data
showed up 8 frequencies of the $\beta$ Cep type and 6 peaks of the SPB type.
Using these observations, we were able to determine the empirical values of $f$
for four $\beta$ Cep modes and for all six SPB modes.
The work on this variable is in the making and results will appear soon.

\section{Conclusions}
Presented examples demonstrate a great potential of \textbf{complex seismic studies}
consisting in fitting simultaneously pulsational frequencies and corresponding values of
the complex, nonadiabatic parameter  $f$.
In particular, complex asteroseismology of the B-type pulsators yields a valuable constraints
on \textbf{stellar opacities}. Heretofore, our results indicate a preference for \textbf{the OPAL tables}.

One of the most important results is that empirical values of $f$
are determinable also for \textbf{high order g modes}. The $f$ parameter of the $\beta$ Cep and SPB modes
have a different dependence on the mode frequency and degree, $\ell$.
Therefore, complex asteroseismology of the \textbf{hybrid pulsators} can give
much better assessment of stellar model precision.

\begin{acknowledgments}
The work was supported by Astronomical Institute of Wroc{\l}aw University
and by the Polish MNiSW grant N N203 379636.
\end{acknowledgments}

\end{document}